\documentclass[conference]{IEEEtran}
\IEEEoverridecommandlockouts
\usepackage{cite}
\usepackage{amsmath,amssymb,amsfonts}
\usepackage{algorithmic}
\usepackage{graphicx}
\usepackage{textcomp}
\usepackage{xcolor}
\usepackage{array}
\usepackage{booktabs} % 提供更专业的表格线条
\usepackage{newtxtext} % 使用新的Times字体，更现代
\usepackage{color}
\usepackage{hyperref}
\usepackage{url}
\usepackage{tabularx}
\usepackage[table]{xcolor} % 用于表格颜色填充
\usepackage{float}
\usepackage{booktabs}
\usepackage{multirow}
\usepackage{cleveref}
\usepackage{mathrsfs} % 加载 mathrsfs 宏包
\usepackage{graphicx} % 用于插入图片
\usepackage{subcaption} % 用于支持子图
\usepackage{makecell}

\usepackage{pifont}
\usepackage{booktabs} % 用于 \Xhline
\renewcommand{\arraystretch}{1.2}
\def\BibTeX{{\rm B\kern-.05em{\sc i\kern-.025em b}\kern-.08em
    T\kern-.1667em\lower.7ex\hbox{E}\kern-.125emX}}
    
\begin{document}

\title{Iterative LLM-Based Assertion Generation Using Syntax–Semantic Representations for Functional Coverage–Guided Verification
\thanks{\textdagger~ Corresponding Author
\\This paper has been published in the 31st IEEE European Test Symposium (ETS 2026), 25-29th May, 2026, Chania, Greece.}
}

\author{
\IEEEauthorblockN{
Yonghao Wang\textsuperscript{1},
Jiaxin Zhou\textsuperscript{3},
Yang Yin\textsuperscript{1},
Hongqin Lyu\textsuperscript{1,2}, 
Zhiteng Chao\textsuperscript{1},
Wenchao
Ding\textsuperscript{4},
\\
Jing
Ye\textsuperscript{1,2},
Tiancheng
Wang\textsuperscript{1\textdagger}
and
Huawei Li\textsuperscript{1,2\textdagger}}
\IEEEauthorblockA{\textsuperscript{1}State Key Lab of Processors, Institute of Computing Technology, CAS, Beijing, China}
\IEEEauthorblockA{\textsuperscript{2}University of Chinese Academy of Sciences, Beijing, China}
\IEEEauthorblockA{\textsuperscript{3}Beijing Normal University, Beijing, China; \textsuperscript{4}Tencent Ltd., Beijing, China}
\IEEEauthorblockA{wangyonghao22@mails.ucas.ac.cn, \{wangtiancheng, lihuawei\}@ict.ac.cn}
}

\maketitle

\begin{abstract}
While leveraging LLMs to automatically generate SystemVerilog assertions (SVAs) from natural language specifications holds great potential, existing techniques face a key challenge: LLMs often lack sufficient understanding of IC design, leading to poor assertion quality in a single pass. Therefore, verifying whether the generated assertions effectively cover the functional specifications and designing feedback mechanisms based on this coverage remain significant hurdles.
To address these limitations, this paper introduces CoverAssert, a novel iterative framework for optimizing SVA generation with LLMs. The core contribution is a lightweight mechanism for matching generated assertions with specific functional descriptions in the specifications. CoverAssert achieves this by clustering the joint representations of semantic features of LLM-generated assertions and structural features extracted from abstract syntax trees (ASTs) about signals related to assertions, and then mapping them back to the specifications to analyze functional coverage quality.
Leveraging this capability, CoverAssert constructs a feedback loop based on functional coverage to guide LLMs in prioritizing uncovered functional points, thereby iteratively improving assertion quality. Experimental evaluations on four open-source designs demonstrate that integrating CoverAssert with state-of-the-art generators, AssertLLM and Spec2Assertion, achieves average improvements of 9.57\% in branch coverage, 9.64\% in statement coverage, and 15.69\% in toggle coverage.
\end{abstract}


\section{Introduction}

Functional verification plays a crucial role in the realm of integrated circuit (IC) design. Verification engineers meticulously evaluate whether the register-transfer level (RTL) code adheres to the prescribed architectural specifications. Assertion-Based Verification (ABV) has garnered widespread adoption in RTL design, primarily due to its efficacy in enhancing visibility and substantially reducing the time required for simulation debugging, with reductions of up to 50\% \cite{1, 2}.
\begin{figure}[h]
  \centering
  \includegraphics[width=1\linewidth]{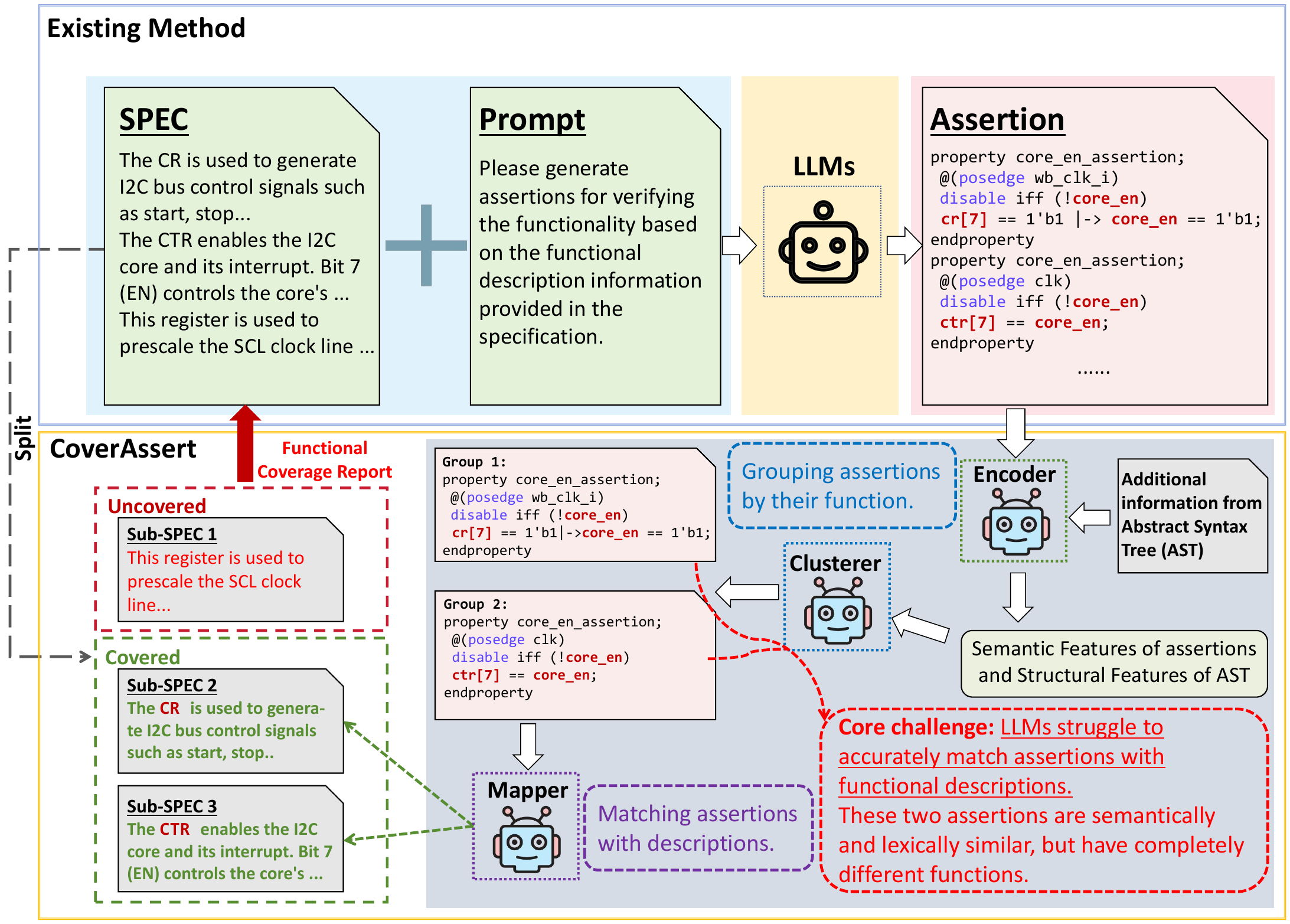}
  \caption{Our proposed framework, CoverAssert, directly enhances assertion generation by accurately identifying uncovered functional descriptions in the specification and providing feedback to prioritize the completion of uncovered functional points, thereby improving overall verification coverage.}
  \label{fig:intro}
\end{figure}
Within the ABV framework, the significance of high-quality SystemVerilog assertions (SVA) for formal property verification (FPV) cannot be overstated \cite{3, 4}.

Ambiguities in specifications \cite{9} and varying designer interpretations make translating specifications into effective SVAs labor-intensive and time-consuming \cite{7,0-16}. With increasing design complexity, efficient SVA generation has become crucial. Recent advances in LLMs have shown great potential to revolutionize this process, offering innovative ways to automate assertion generation. These methods fall into two primary approaches: pre-RTL SVA generation, which directly converts natural language specifications into assertions \cite{10,11,14}, and methods that integrate both specifications and RTL code to capture finer functional and structural details \cite{6,12,21,0-15}.

Despite recent advancements in automated assertion generation by LLMs, significant challenges persist. Current frameworks fail to fully cover the functional descriptions in specifications, creating a critical bottleneck. Without identifying uncovered points, LLMs struggle to self-correct, leaving blind spots in some areas while generating redundant assertions in others, reducing testing efficiency. This issue arises from insufficient domain-specific training for SVA and the sheer complexity of specifications, which makes omissions inevitable.

Therefore, it is essential to develop an effective mechanism that can accurately identify functional intent—that is, the specific functionality or property under verification—and signal relationships of assertions, prompting the precise filtering of uncovered functional descriptions in the specification for functional coverage feedback, thereby effectively guiding the generation process. 
However, matching assertions to functional descriptions is particularly challenging due to the high similarity of assertion code, which often leads LLMs to conflate semantic meaning with lexical similarity as shown in Fig. \ref{fig:intro}. This issue is further exacerbated by overlapping signal names and similar functions across different modules. As a result, accurately aligning assertions with specification statements—and introducing a lightweight coverage feedback mechanism to prioritize uncovered functional points—remains a significant open challenge.

To address these challenges, we propose a lightweight functional coverage feedback loop framework named CoverAssert that directs existing LLM-based SVA generators to refine generated results. For the generated assertions, we encode both syntactic and semantic features, perform feature fusion, and then apply clustering to group assertions. Simultaneously, the original functional specification is divided into sub-specification chunks according to module division. Each assertion group is matched to the most relevant sub-specification chunk and further match the specific functional description statements within them, allowing us to evaluate the functional coverage of the generated assertions. Uncovered sub-specification chunks are then fed back iteratively to improve the assertion generation quality. The overall framework, illustrated in Fig. \ref{fig:intro}, significantly alleviates the bottleneck where LLMs struggle to analyze functional coverage for assertions with highly similar syntax and semantics.

The contributions of this paper are summarized as follows:

\begin{enumerate}[]
\item We propose CoverAssert, a lightweight LLM SVA iterative generation framework. It analyzes the coverage of functional specifications by extracting the syntax-semantic representations of the currently generated assertions. To the best of our knowledge, this is the first work to propose an LLM assertion generation paradigm based on functional coverage feedback, which can be seamlessly integrated with any existing method.

\item In addition to using LLM embeddings to encode assertion semantic features, CoverAssert analyzes the relative positional relationships of variables in the syntax tree and introduces structural features, thereby enabling the feature fusion representation of assertions to precisely match the functional descriptions in the specification.

\item When integrated with two state-of-the-art methods, experiments on four open-source circuit cases demonstrate that CoverAssert achieves average improvements of 9.57\% in branch coverage, 9.64\% in statement coverage, and 15.69\% in toggle coverage.
\end{enumerate}

\section{Background}

\subsection{Assertion Generation Based on LLM}

The early work of automated hardware assertion generation proposed by Rahul Kande et al. \cite{12}, who pioneered the application of LLMs to generate assertions. AssertLLM by Fang et al. \cite{10} was a milestone, handling full specification files to produce detailed SVAs for all architectural signals. Bai et al. \cite{6} contributed AssertionForge, which built a unified Knowledge Graph (KG) from both natural language specs and RTL code, linking high-level design intent with low-level implementation details. More recently, Wu et al. \cite{14} introduced Spec2Assertion, using progressive regularization and Chain-of-Thought prompting to generate high-quality assertions directly from specification.

\subsection{Semantic Similarity Computation}

Semantic similarity computation measures the degree of semantic equivalence between two textual or structural artifacts by mapping inputs to dense vector representations and deriving similarity scores, often using cosine similarity \cite{1-5}. The advent of distributed representations, such as word2vec \cite{1-6}, enabled context-aware similarity at the token level. Sentence-level techniques used recurrent or convolutional networks to aggregate word embeddings into holistic representations. Transformer-based models like BERT \cite{1-9} further advanced the field by producing contextualized embeddings that capture long-range dependencies and semantic nuances.

\section{Framework of CoverAssert}

\begin{figure*}[h]
  \centering
  \includegraphics[width=0.87\linewidth]{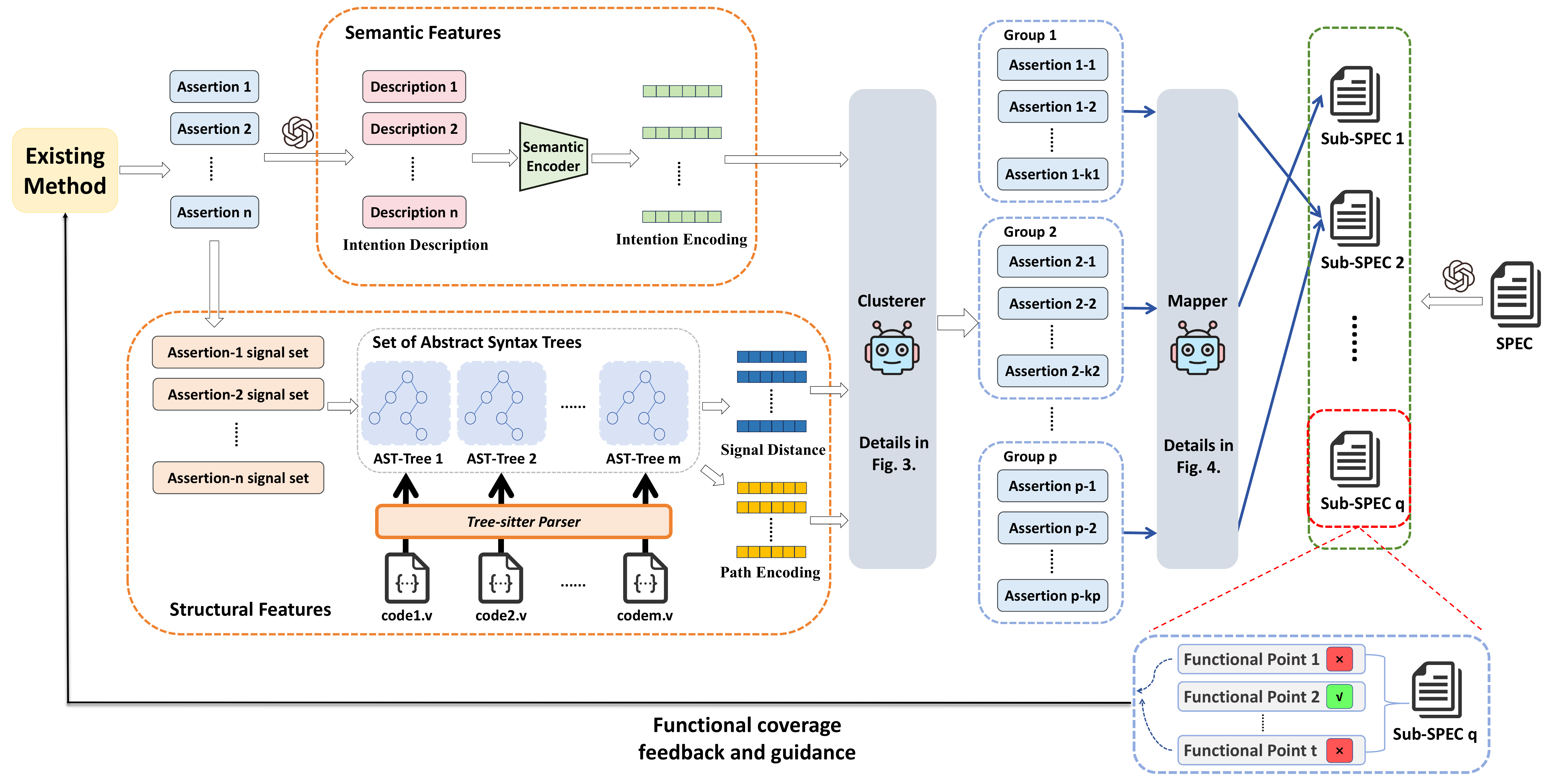}
  \caption{CoverAssert integrates with existing assertion generation methods by extracting semantic and structural features of assertions and signals, which are fused and clustered, then matched with segmented Sub-SPECs. Insufficiently covered Sub-SPECs and their functional descriptions are fed back to guide targeted assertion generation, thereby improving coverage.}
  \label{fig:framework}
\end{figure*}

\subsection{Workflow Overview}

This paper proposes CoverAssert, a functional coverage–guided iterative framework that leverages syntax–semantic dual-modal clustering to identify coverage gaps and guide targeted assertion generation. The six-module workflow is shown in Fig. \ref{fig:framework}.

\subsection{Semantic Feature Extraction Module}

Due to high syntactic similarity, directly embedding assertion code fails to capture semantic distinctions. We therefore use an LLM (e.g., ChatGPT-4o) to extract each assertion’s intent as a concise natural-language description, which is then encoded by Qwen3-Embedding into compact semantic vectors. This process yields semantically discriminative representations and forms the final vector matrix.
\begin{equation}
\mathbf{T} \in \mathbb{R}^{N \times 4096}
\end{equation}
where $N$ denotes the number of assertions, and 4096 is the output dimension of the model that we adopted to preserve richer semantic information.

\subsection{Signal Structural Features Extraction Module}

The extreme syntactic similarity of assertions causes LLMs to conflate semantic meaning with lexical similarity when generating textual intents. To eliminate this ambiguity, we explicitly incorporate structural features inherent to each assertion from the ASTs. Formally, for every assertion $a_i$, we first extract the referenced signals into a set $S_i$. Subsequently, the original RTL files are parsed into multiple ASTs by leveraging \textit{Tree-sitter} \cite{0-13}, where each AST retains fully-qualified hierarchical identifiers as keys:

\begin{equation}
\mathcal{N}:{signal\_name\rightarrow\{Node\}}
\end{equation}
For every signal pair $(s, t) \in S_i \times S_j$ (across assertions $a_i$ and $a_j$), we compute the lowest common ancestor (LCA) as:

\begin{equation}
\begin{split}
    d(s, t) = \bigl|depth(s) - depth\bigl(LCA(s,t)\bigr)\bigr| \\
    \quad + \bigl|depth(t) - depth\bigl(LCA(s,t)\bigr)\bigr|
\end{split}
\end{equation}
where $LCA(s, t)$ denotes the lowest common ancestor of nodes $s$ and $t$ within the same AST, if two signals are located in different ASTs, the maximum penalty will be imposed. 

For any assertion pair $(a_i, a_j)$, the final LCA-based structural distance is defined as the average of the minimal pairwise distances over their signal sets:
\begin{equation}
SD_{ij}=\frac{1}{|S_i||{S_j}|}\sum_{v \in S_i}^{} \sum_{u \in S_j}^{}\min\limits_{\substack{n_v \in \mathcal{N}(v) \\ n_u \in \mathcal{N}(u)}} d(n_v, n_u)
\end{equation}
Thus, we capture the structural distance information between the signals of each assertion pair as the first structural feature, and construct a symmetric assertion structural distance matrix:
\begin{equation}
\mathbf{SD} \in \mathbb{R}^{N \times N}
\end{equation}

In addition to this, for each assertion, the positional information of each signal within the AST is extracted, and these pieces of information are concatenated to form a comprehensive feature vector.   Suppose there are $n$ signals $v_1,v_2,...,v_n$, and the path of each signal $v_i$ in the AST is denoted as $Path(v_i)$, which can be represented as:
\begin{equation}
Path(v_i)=[n_0^{(i)}, n_1^{(i)},...n_{d_i}^{(i)}]
\end{equation}
where $n_0^{(i)}$ is the root node, $n_{d_i}^{(i)}$ is the node where signal $v_i$ is located, and $d_i$ is the depth of signal $v_i$. After sorting the signal names in $S_i$, the paths of all signals are concatenated into a comprehensive feature vector $Q_i$:
\begin{equation}
Q_i = Path(s_1) \oplus Path(s_2) \oplus \cdots \oplus Path(s_n)
\end{equation}

To standardize the dimensions of feature vectors $Q_i$ derived from assertions with varying numbers of signals, we pad shorter vectors with zeros to match the maximum dimension $D_{max}$. This results in a matrix $\mathbf{Q}$ as the second structural feature:
\begin{equation}
\mathbf{Q} \in \mathbb{R}^{N \times D_{max}}
\end{equation}

\subsection{Clustering Module}

\begin{figure}[h]
\centering
\includegraphics[width=0.85\linewidth]{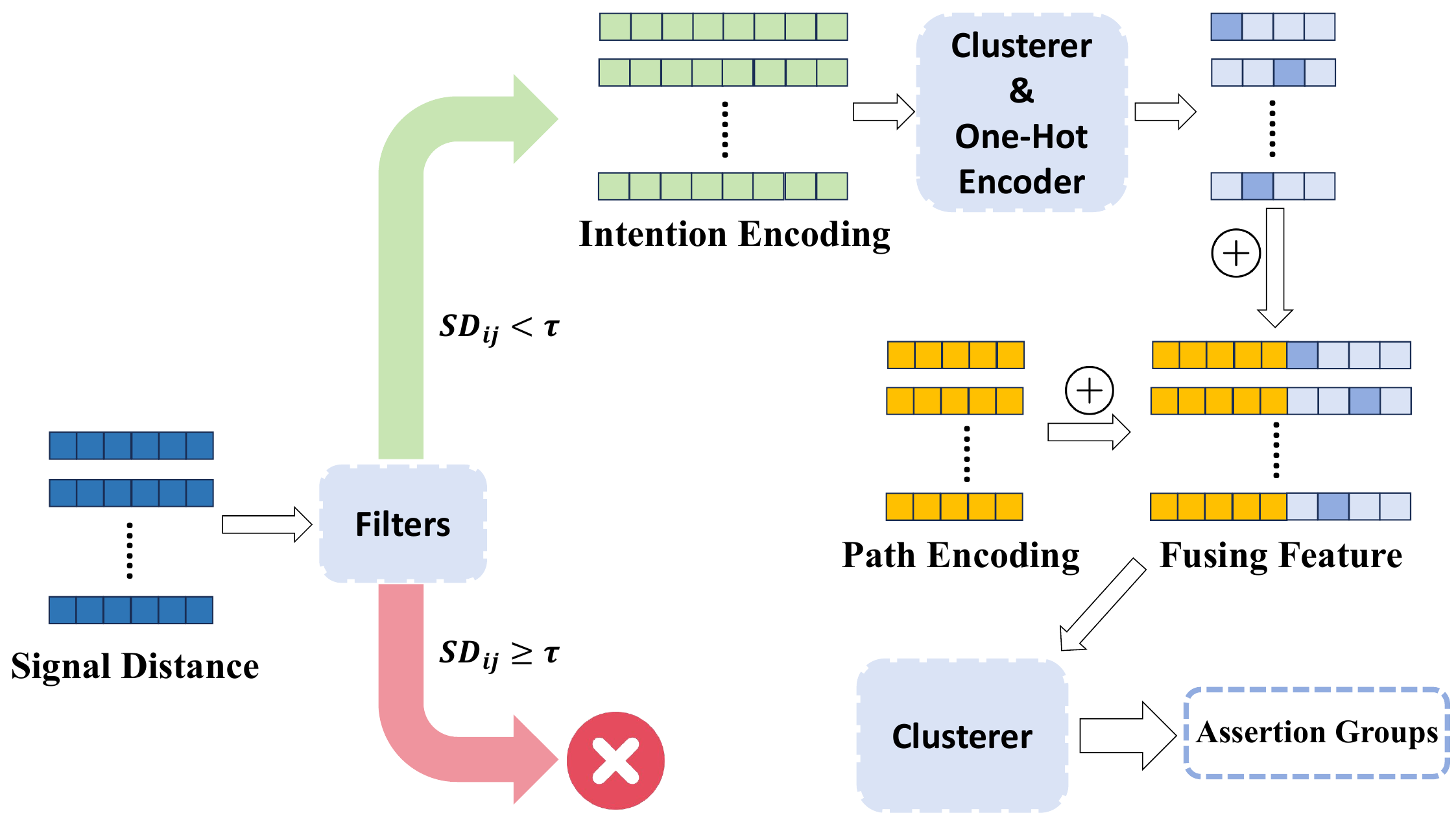}
\caption{Flowchart of clustering process via fusion of structural and semantic features of assertions.}
\label{feature_fusion}
\end{figure}

To address high assertion similarity and overlapping signal names, we group assertions modularly based on extracted features, enabling preliminary matching with the specification. To further improve feature discriminability and clustering accuracy, we apply a feature fusion and clustering approach, as shown in Fig. \ref{feature_fusion}.

We exclude assertion pairs with $SD_{ij}>\tau ~(\tau=15)$ and cluster only those with $SD_{ij}\leq\tau$. To address the disparity between semantic and structural features, we first perform preliminary clustering based on semantic features. Then, we apply the \texttt{DBSCAN} clustering algorithm \cite{0-12} to $\mathbf{T}$ to obtain cluster labels $C_i$ for each assertion. These labels serve as a dimensionality-reduced representation of semantic features. We encode each cluster label $C_i$ into a one-hot vector and concatenate it with the normalized and PCA-reduced structural feature vector $PCA(\hat{Q_i})$ to form the fused feature vector $Q_i'$:
\begin{equation}
Q_i' = PCA(\hat{Q_i}) \oplus one\text{-}hot(C_i)
\end{equation}
where $PCA(\hat{Q_i})$ is a 20-dimensional PCA projection retaining an \texttt{Explained Variance Ratio} \cite{0-14} above 0.97, yielding the fused feature matrix:
\begin{equation}
\mathbf{Q}' \in \mathbb{R}^{N \times (20 + K)}
\end{equation}
where $K$ is the dimension of $one\text{-}hot(C_i)$. The fused feature matrix $\mathbf{Q}'$ is then clustered, with Silhouette Score \cite{0-11} used to evaluate and adjust clustering. Assertions are ultimately divided into multiple groups.

\subsection{Specification Split and Functional Points Extraction Module}

To match assertions with concrete descriptions in the original specification, we use LLM to extract functional modules as Sub-SPECs, each containing signal names, functional descriptions, and other details. We then extract fine-grained \textbf{validation functional points—concise, atomic statements that must be verified directly from the specification}. An example extracted from a Sub-SPEC is:
\begin{equation*}
\scriptsize
\begin{aligned}
&\texttt{\textbf{SPEC q-1}: $go$ should be high when any of the commands ($start$,} \\
&\texttt{$stop$, $read$, $write$) are asserted, and $cmd\_ack$ is low.}
\end{aligned}
\end{equation*}
This allows each Sub-SPEC to cover a full function module with many fine-grained functional points, enabling more precise feedback in subsequent stages.

\subsection{Assertion-to-Specification Mapping Module}

\begin{figure}[h]
\centering
\includegraphics[width=0.85\linewidth]{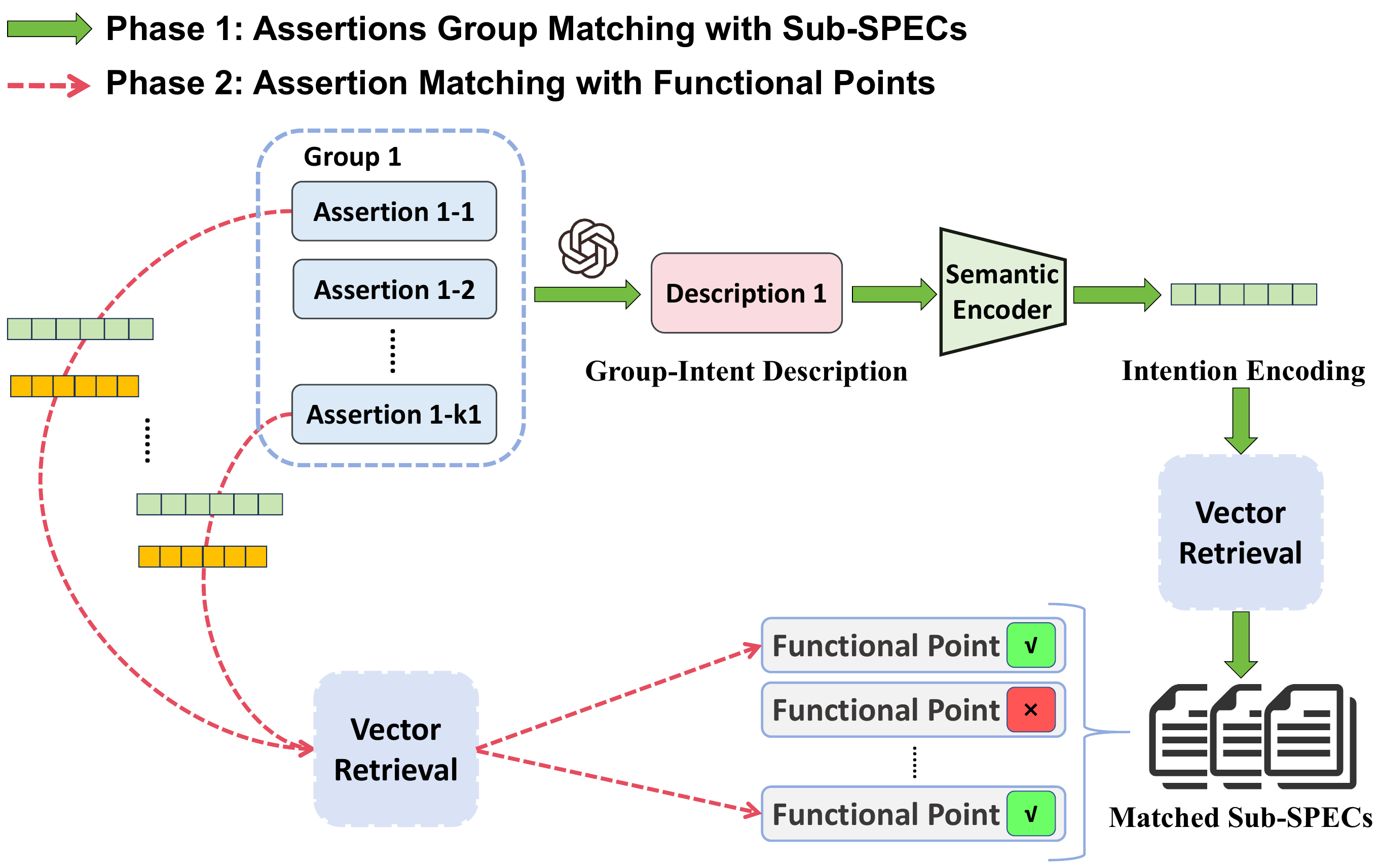}
\caption{Flowchart of the matching process between assertions, Sub-SPECs, and functional points.}
\label{spec_match}
\end{figure}

To prevent conflating similar intents, assertion groups are first modularly matched to the most relevant Sub-SPEC according to clustering results, ensuring each group corresponds to the correct functional module. Each assertion is then aligned with the specific functional points within the assigned Sub-SPEC, as illustrated in Fig. \ref{spec_match}.

We first use LLM to generate functional descriptions with associated signals for each assertion group, reducing ambiguity from similar syntactic structures. Vector retrieval identifies inadequately covered Sub-SPECs by comparing them with group intents, after which signal names and functional descriptions are used to match each assertion to specific functional points. This ensures precise alignment, and any uncovered Sub-SPECs and their points are sent to the Coverage-Driven Feedback Loop Module.

\subsection{Coverage-Driven Feedback Loop Module}

The assertion-to-specification mapping module identifies uncovered Sub-SPECs and their validation points, which are fed back to guide the assertion generation engine. The coverage-driven loop focuses on under-constrained regions, iteratively generating assertions for uncovered points and improving functional coverage. The process terminates automatically when each Sub-SPEC’s match degree exceeds the threshold $\theta=0.85$.

\section{Experiments}

\subsection{Experimental Setup}

The benchmark dataset \cite{18} comprises 20 designs, each with a specification file and corresponding golden RTL code. Correctness was analyzed using Cadence JasperGold (v21.12.002). Experiments ran on a server with an Intel Xeon Gold 6148 CPU at 2.40GHz. To evaluate CoverAssert, we conducted comparative experiments against two SOTA assertion generation methods under uniform conditions, with all three using GPT-4o as the experimental model.

\begin{itemize}
    \item \textbf{AssertLLM \cite{10}:} This method uses a multi-agent system to iteratively extract signal descriptions from specification documents, maps them to RTL signals, and employs LLMs to generate assertions, enabling assertion creation before RTL implementation.
    \item \textbf{Spec2Assertion \cite{14}:} This method utilizes LLMs equipped with progressive regularization and Chain-of-Thought prompting to formalize and extract functional descriptions, thereby directly generating pre-RTL assertions from design specifications.
\end{itemize}

\subsection{Evaluation Metrics and Benchmarks}

We use the metrics in JasperGold \cite{19}, with evaluation criteria in Table \ref{Summary of Evaluation Metrics}.

\begin{table}[h]
\centering
\renewcommand{\arraystretch}{1} % 调整行间距
\caption{Summary of evaluation metrics.}
\label{Summary of Evaluation Metrics}
\begin{tabular}{c|c}
\hline
\hline
\cellcolor{gray!20}\textbf{Metrics} & \cellcolor{gray!20}\textbf{Summary} \\ 
\hline
\makecell{$N$} & \makecell{\hspace*{-25pt} The number of generated SVAs}\\ 
\makecell{$S$} & \makecell{\hspace*{-10pt} The number of syntax-correct SVAs}\\
\makecell{$P$} & \makecell{\hspace*{-18pt} The number of FPV-passed SVAs}\\
\makecell{\textbf{$BFC$}} & \makecell{\hspace*{-14pt} Branch coverage in formal analysis} \\ 
\makecell{\textbf{$SFC$}} & \makecell{\hspace*{-5pt} Statement coverage in formal analysis} \\ 
\makecell{\textbf{$TFC$}} & \makecell{\hspace*{-15pt} Toggle coverage in formal analysis} \\
\hline
\hline
\end{tabular}
\\
\vspace{0.1cm} % 添加垂直间距
\scriptsize % 更小的字体
\raggedright % 设置文本左对齐
*Note: The coverage metrics \textbf{$TFC$} are computed within the \textbf{Cones of Influence (COI)} \cite{19} covered by the generated assertions, where a larger \textbf{COI} encompasses more signals. Therefore, in each design experiment, we use the number of signals involved in the largest \textbf{COI} as the standard denominator for the calculation.
\end{table}

To provide an intuitive assessment of CoverAssert, we selected four designs of varying scales: I2C, ECG, Pairing, and SHA3. Detailed descriptions are in Table \ref{Summary of Design}, where LoC denotes lines of code after synthesis.

\begin{table}[h]
\centering
\renewcommand{\arraystretch}{1} % 调整行间距
\caption{Summary of designs.}
\label{Summary of Design}
\setlength{\tabcolsep}{3.5pt}
\begin{tabular}{c|c|c|c}
\hline
\hline
\cellcolor{gray!20}\textbf{Design Name} & \cellcolor{gray!20}\textbf{Func. Description} & \cellcolor{gray!20}\textbf{LoC} &
\cellcolor{gray!20}\textbf{Num. of Cells} \\ \hline
\makecell{\texttt{I\textsuperscript{2}C}} & \makecell{\hspace*{-4pt} Serial communication protocol.} & \makecell{5369} & \makecell{756}\\
\makecell{\texttt{SHA3}} & \makecell{\hspace*{-17pt} Hash function computation.} & \makecell{141185} & \makecell{22228}\\ 
\makecell{\texttt{ECG}} & \makecell{\hspace*{-12pt} Biological signal acquisition.} & \makecell{398686} & \makecell{59084}\\ 
\makecell{\texttt{Pairing}} & \makecell{\hspace*{-13pt} Cryptographic key exchange.} & \makecell{1561498} & \makecell{228287}\\ 
\hline
\hline
\end{tabular}
\end{table}

\subsection{Experimental Results}

\subsubsection{\textbf{Integration Experiments of CoverAssert with Previous Methods}}

To validate CoverAssert's efficacy, we integrated it with AssertLLM and Spec2Assertion to observe coverage changes across four circuits, as shown in Table \ref{tab:assert+cover}. $CoverAssert\textbf{-}n$ denotes the state after n iterations of the feedback loop. Most circuits automatically terminate after two iterations, so results are only shown for two. 

\definecolor{deepgreen}{RGB}{110, 251, 152}
\definecolor{lightgreen}{RGB}{188, 251, 152}

\begin{table}[h]
\centering
\caption{Performance comparison of integrating CoverAssert with AssertLLM and Spec2Assertion.}
\label{tab:assert+cover}
\setlength{\tabcolsep}{0.4em} % 调整列间距
\renewcommand{\arraystretch}{0.6} % 缩小行间距
\begin{tabular}{@{}>{\setlength{\tabcolsep}{-2em}}l *{5}{c} @{}}
\toprule
\hline
\multicolumn{1}{c}{\cellcolor{gray!20}\textbf{Method}} & \cellcolor{gray!20}\textbf{Metrics} & \cellcolor{gray!20}\textbf{I\textsuperscript{2}C} & \cellcolor{gray!20}\textbf{SHA3} & \cellcolor{gray!20}\textbf{ECG} & \cellcolor{gray!20}\textbf{Pairing} \\
\midrule
\multirow{5}{*}{$\mathrm{AssertLLM}$} & $N$/$S$/$P$ & 127/112/58 & 31/31/26 & 44/44/19 & 32/32/12 \\
 & $BFC(\%)$ & 80.23 & 92 & 82.22 & 76.12 \\
 & $SFC(\%)$ & 82.26 & 90.24 & 80.74 & 83.63 \\ 
 & $TFC(\%)$ & 62 & 78.41 & 57.89 & 67.12 \\
\addlinespace
\multirow{5}{*}{\begin{tabular}{@{}l@{}}$\mathrm{AssertLLM}$ \\ \ \ \ \ \ \ \ $\mathrm{+}$ \\ $\mathrm{CoverAssert\text{-}1}$\end{tabular}} & $N$/$S$/$P$ & 149/133/76 & 69/67/47 & 62/60/31 & 51/49/26 \\
 & $BFC(\%)$ & \cellcolor{lightgreen}85.33 & \cellcolor{lightgreen}98.82 & \cellcolor{lightgreen}98.89 & \cellcolor{lightgreen}90.96 \\
 & $SFC(\%)$ & \cellcolor{lightgreen}87.27 & \cellcolor{lightgreen}94.63 & \cellcolor{lightgreen}97.78 & \cellcolor{lightgreen}93.18 \\ 
 & $TFC(\%)$ & \cellcolor{lightgreen}66.98 & \cellcolor{lightgreen}86.57 & \cellcolor{lightgreen}81.86 & \cellcolor{lightgreen}70.33 \\
\addlinespace
\multirow{5}{*}{\begin{tabular}{@{}l@{}}$\mathrm{AssertLLM}$ \\ \ \ \ \ \ \ \ $\mathrm{+}$ \\ $\mathrm{CoverAssert\text{-}2}$\end{tabular}} & $N$/$S$/$P$ & 173/151/85 & 82/80/56 & 78/76/38 & 64/62/32 \\
 & $BFC(\%)$ & \cellcolor{deepgreen}86.79 & \cellcolor{deepgreen}100 & \cellcolor{lightgreen}98.89 & \cellcolor{deepgreen}93.85 \\
 & $SFC(\%)$ & \cellcolor{deepgreen}88.87 & \cellcolor{deepgreen}95.06 & \cellcolor{lightgreen}97.78 & \cellcolor{deepgreen}94.67 \\ 
 & $TFC(\%)$ & \cellcolor{deepgreen}80.52 & \cellcolor{deepgreen}90.71 & \cellcolor{deepgreen}83.27 & \cellcolor{deepgreen}81.46 \\
\addlinespace
\hline
\hline
\\
\multirow{5}{*}{$\mathrm{Spec2Assertion}$} & $N$/$S$/$P$ & 90/89/49 & 42/42/28 & 35/29/19 & 45/41/15 \\
 & $BFC(\%)$ & 87.87 & 90.89 & 79.51 & 83.58 \\
 & $SFC(\%)$ & 89.44 & 87.78 & 81.05 & 66.67 \\ 
 & $TFC(\%)$ & 64.13 & 65.43 & 73.92 & 46.53 \\
\addlinespace
\multirow{5}{*}{\begin{tabular}{@{}l@{}}$\mathrm{Spec2Assertion}$ \\ \ \ \ \ \ \ \ $\mathrm{+}$ \\ $\mathrm{CoverAssert\text{-}1}$\end{tabular}} & $N$/$S$/$P$ & 105/104/61 & 65/65/46 & 51/42/29 & 76/68/34 \\
 & $BFC(\%)$ & \cellcolor{lightgreen}89.34 & \cellcolor{lightgreen}94.36 & \cellcolor{lightgreen}97.33 & \cellcolor{lightgreen}86.91 \\
 & $SFC(\%)$ & \cellcolor{lightgreen}90.32 & \cellcolor{lightgreen}90.84 & \cellcolor{lightgreen}96.48 & \cellcolor{lightgreen}81.06 \\ 
 & $TFC(\%)$ & \cellcolor{lightgreen}75.06 & \cellcolor{lightgreen}74.03 & \cellcolor{lightgreen}75.26 & \cellcolor{lightgreen}58.35 \\
 \addlinespace
\multirow{5}{*}{\begin{tabular}{@{}l@{}}$\mathrm{Spec2Assertion}$ \\ \ \ \ \ \ \ \ $\mathrm{+}$ \\ $\mathrm{CoverAssert\text{-}2}$\end{tabular}} & $N$/$S$/$P$ & 117/116/67 & 77/77/55 & 68/58/37 & 101/90/51 \\
 & $BFC(\%)$ & \cellcolor{lightgreen} 89.34 & \cellcolor{deepgreen}95.07 & \cellcolor{deepgreen} 98.02 & \cellcolor{lightgreen} 86.91 \\
 & $SFC(\%)$ & \cellcolor{lightgreen} 90.32 & \cellcolor{deepgreen}92.84 & \cellcolor{lightgreen}96.48 & \cellcolor{deepgreen}82.94 \\ 
 & $TFC(\%)$ & \cellcolor{deepgreen}82.42 & \cellcolor{deepgreen}77.75 & \cellcolor{deepgreen}82.85 & \cellcolor{deepgreen}61.96 \\
\hline
\bottomrule
\end{tabular}
\end{table}

As the results show, the first iteration significantly boosts all coverage metrics, and the second iteration further improves most metrics. Each iteration adds only a few assertions, demonstrating precise and effective guidance.

\subsubsection{\textbf{Analysis of the Impact of Feedback Iteration on the Quality of Assertion Generation}}

\begin{figure}[h]
\centering
\includegraphics[width=1\linewidth]{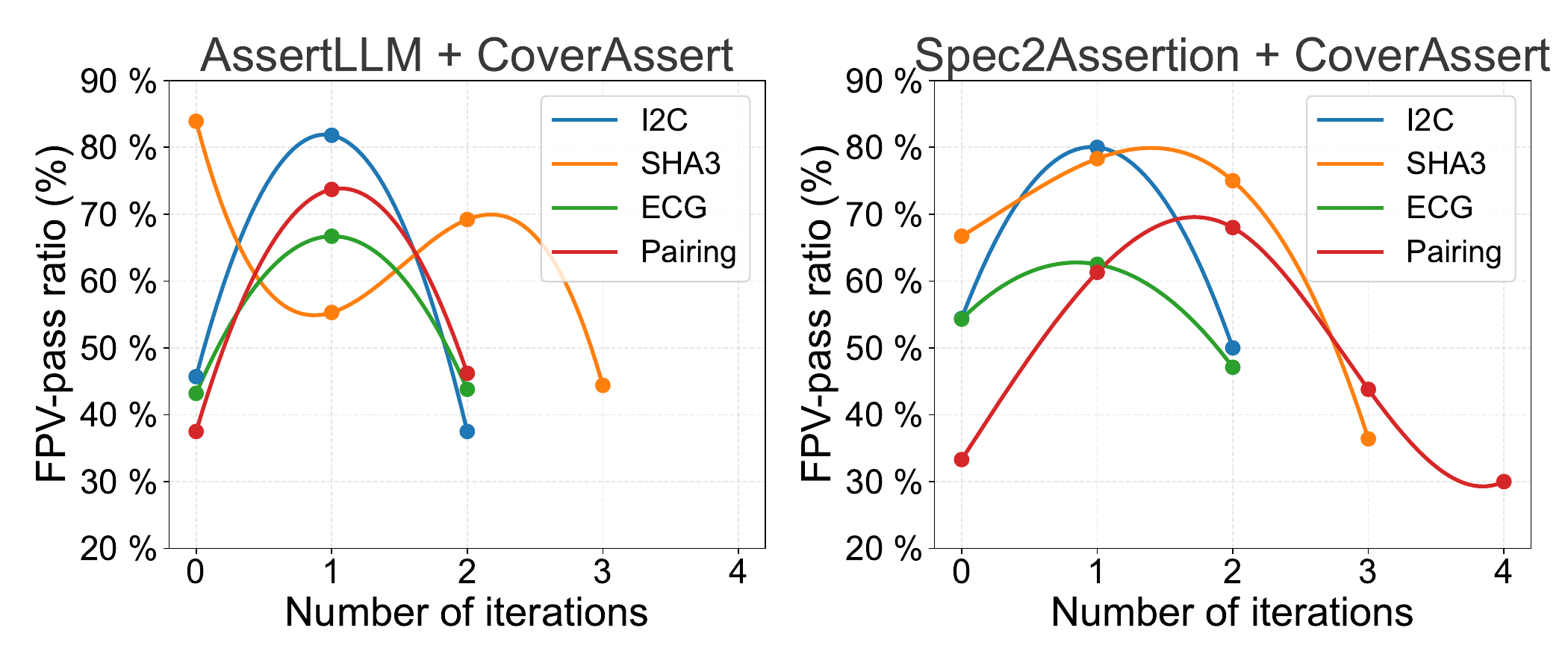}
\caption{FPV pass-rate evolution across feedback iterations.}
\label{fig:iteration}
\end{figure}

To evaluate the feedback loop’s impact, we monitored the FPV-pass rate of assertions in each iteration, and compared it with the rate of the original non-iterative algorithm, as shown in Fig. \ref{fig:iteration}.

Experiments show FPV-pass curves are mostly convex: precision rises sharply in early iterations as many uncovered features are found, then declines in later iterations as the generator tackles harder, vaguely specified functions. This confirms that the feedback loop focuses generation on difficult portions. For SHA3, its limited internal iteration structure allows easy functional points to be covered early, while remaining hard-to-reach points cause a larger drop in iteration accuracy.

\subsubsection{\textbf{Ablation Study - Removing Signal Structural Features \& Replace All Components with ChatGPT}}

We perform two ablations to assess structural features and clustering: (1) retaining only semantic embeddings, and (2) letting ChatGPT-4o fully handle feature extraction and assertion–functional description matching. One-iteration results are in Table \ref{tab:assert+cover+wo}.

\begin{table}[h]
\centering
\caption{Performance comparison of integrating CoverAssert (CA) w/o structural features and ChatGPT-4o with AssertLLM and Spec2Assertion.}
\label{tab:assert+cover+wo}
\setlength{\tabcolsep}{0.3em} % 调整列间距
\renewcommand{\arraystretch}{0.6} % 缩小行间距
\begin{tabular}{@{}>{\setlength{\tabcolsep}{-2em}}l *{5}{c} @{}}
\toprule
\hline
\multicolumn{1}{c}{\cellcolor{gray!20}\textbf{Method}} & \cellcolor{gray!20}\textbf{Metrics} & \cellcolor{gray!20}\textbf{I\textsuperscript{2}C} & \cellcolor{gray!20}\textbf{SHA3} & \cellcolor{gray!20}\textbf{ECG} & \cellcolor{gray!20}\textbf{Pairing} \\
\midrule
\multirow{5}{*}{\begin{tabular}{@{}l@{}}$\mathrm{AssertLLM}$ \\ \ \ \ \ \ \ \ $\mathrm{+}$ \\ $\mathrm{CA\text{-}1}$ w/o struc. Feat.\end{tabular}} & $N$/$S$/$P$ & 138/122/65 & 37/37/29 & 56/54/25 & 52/50/17 \\
 & $BFC(\%)$ & \cellcolor{lightgreen}82.65 & \cellcolor{lightgreen}92.81 & \cellcolor{lightgreen}86.3 & \cellcolor{lightgreen}82.93 \\
 & $SFC(\%)$ & \cellcolor{lightgreen}83.35 & \cellcolor{lightgreen}90.35 & \cellcolor{lightgreen}83.49 & 83.63 \\ 
 & $TFC(\%)$ & \cellcolor{lightgreen}63.34 & 78.41 & \cellcolor{lightgreen}69.13 & 67.12 \\
\addlinespace
\multirow{5}{*}{\begin{tabular}{@{}l@{}}$\mathrm{AssertLLM}$ \\ \ \ \ \ \ \ \ $\mathrm{+}$ \\ $\mathrm{ChatGPT\textit{-}4o\textit{-}1}$\end{tabular}} & $N$/$S$/$P$ & 174/154/87 & 88/88/69 & 101/97/49 & 74/67/31 \\
 & $BFC(\%)$ & \cellcolor{deepgreen}85.71 & \cellcolor{lightgreen}98.51 & \cellcolor{lightgreen}93.61 & \cellcolor{lightgreen}88.61 \\
 & $SFC(\%)$ & \cellcolor{lightgreen}86.91 & \cellcolor{deepgreen}96.6 & \cellcolor{lightgreen}91.92 & \cellcolor{lightgreen}90.91 \\ 
 & $TFC(\%)$ & \cellcolor{deepgreen}73.34 & \cellcolor{deepgreen}88.62 & \cellcolor{lightgreen}74.34 & \cellcolor{lightgreen}68.64 \\
\addlinespace
\hline
\hline
\\
\multirow{5}{*}{\begin{tabular}{@{}l@{}}$\mathrm{Spec2Assertion}$ \\ \ \ \ \ \ \ \ $\mathrm{+}$ \\ $\mathrm{CA\text{-}1}$ w/o struc. Feat.\end{tabular}}  & $N$/$S$/$P$ & 112/108/67 & 48/48/31 & 54/44/31 & 69/65/21 \\
 & $BFC(\%)$ & \cellcolor{lightgreen}87.93 & 90.89 & \cellcolor{lightgreen}90.92 & 83.58 \\
 & $SFC(\%)$ & 89.44 & 87.78 & \cellcolor{lightgreen}88.3 & \cellcolor{lightgreen}69.16 \\ 
 & $TFC(\%)$ & \cellcolor{lightgreen}64.6 & \cellcolor{lightgreen}66.02 & \cellcolor{deepgreen}76.61 & \cellcolor{lightgreen}48.9 \\
\addlinespace
\multirow{5}{*}{\begin{tabular}{@{}l@{}}$\mathrm{Spec2Assertion}$ \\ \ \ \ \ \ \ \ $\mathrm{+}$ \\ $\mathrm{ChatGPT\textit{-}4o\text{-}1}$\end{tabular}} & $N$/$S$/$P$ & 129/128/78 & 74/74/55 & 72/59/41 & 89/81/39 \\
 & $BFC(\%)$ & \cellcolor{deepgreen}90.31 & \cellcolor{lightgreen}90.01 & \cellcolor{lightgreen}91.67 & \cellcolor{lightgreen}83.91 \\
 & $SFC(\%)$ & \cellcolor{lightgreen}89.15 & \cellcolor{deepgreen}93.51 & \cellcolor{lightgreen}89.16 & \cellcolor{lightgreen}70.36 \\ 
 & $TFC(\%)$ & \cellcolor{deepgreen}79.31 & \cellcolor{lightgreen}70.01 & \cellcolor{deepgreen}78.81 & \cellcolor{lightgreen}51.11 \\
 \addlinespace
\hline
\bottomrule
\end{tabular}
\end{table}

The results show that most metrics still improve after one round of feedback, demonstrating the effectiveness of our framework. However, without structural features, the framework either fails to provide feedback or stops after one iteration, yielding minor coverage gains, far behind CoverAssert-1's results in Table \ref{tab:assert+cover}. In contrast, ChatGPT-4o generates more assertions, which improves some metrics for small circuits but performs poorly in large-scale circuits.

\begin{figure}[h]
\centering
\includegraphics[width=1\linewidth]{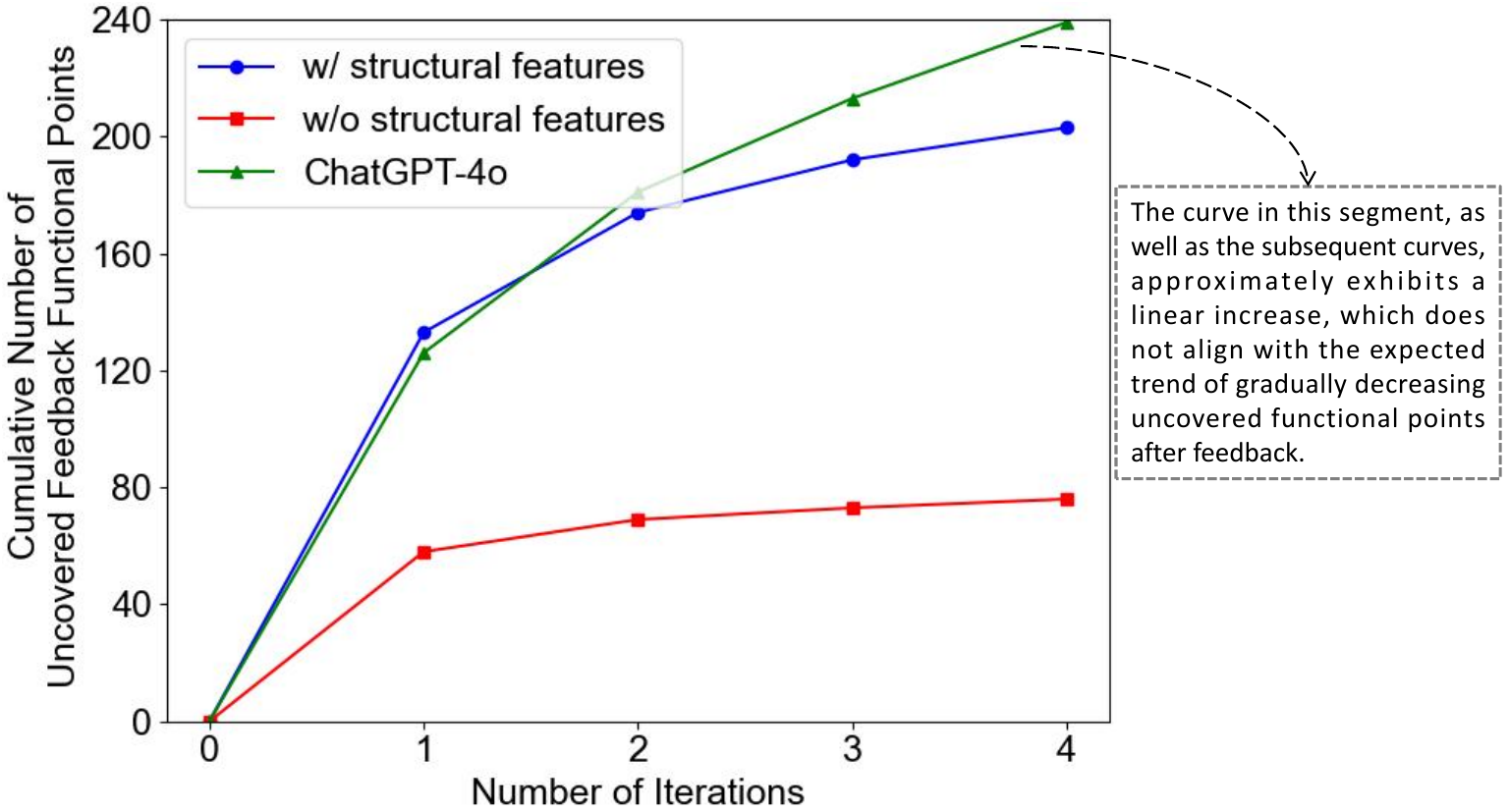}
\caption{Comparison of the iterative feedback cumulative numbers of uncovered functional points for coverAssert with structural features, coverAssert without structural features, and ChatGPT-4o for extracting features and mapping.}
\label{fig:ablation1}
\end{figure}

To analyze this, we increase the Sub-SPEC threshold $\theta$ to allow more iterations, and compare uncovered validation points per iteration across the three methods for four circuits, as shown in Fig. \ref{fig:ablation1}.

Without structural features, assertions often match unrelated descriptions, inflating Sub-SPEC matches and limiting coverage, highlighting the need for structural features and modular grouping.  In addition, ChatGPT may re-output previously covered statements or even produce more uncovered points than earlier iterations.  It tends to generate random feedback when handling complex or indistinguishable statements and thereby reducing the pertinence of its responses.

\section{Conclusion}

This paper presents CoverAssert, a framework that optimizes SystemVerilog assertion generation via functional coverage feedback. By combining semantic and structural features, it improves grouping and matching of assertions with verification targets. Integrated with existing RTL SVA methods, CoverAssert guides generators to prioritize uncovered points, enhancing assertion quality and coverage.

\bibliographystyle{plain} % 使用 plain 样式
% \clearpage % 开始新的一页
\bibliography{reference}

\end{document}